\RequirePackage{fix-cm}
\documentclass{svjour3}                     
\smartqed  
\usepackage{graphicx}
\usepackage[center]{caption}
\usepackage{enumitem}
\usepackage{adjustbox}
\usepackage{diagbox}
\usepackage{booktabs,siunitx}
\usepackage[table,xcdraw]{xcolor}
\usepackage{natbib}
\usepackage{comment}
\journalname{Empirical Software Engineering}
\begin{document}

\title{Crowdsmelling: The use of collective knowledge in code smells detection
}


\author{José Pereira dos Reis \and Fernando Brito e Abreu \and Glauco de Figueiredo Carneiro}


\institute{José Pereira dos Reis \at
              ISTAR-Iscte, Instituto Universitário de Lisboa, Lisboa, Portugal\\
              \email{jvprs@iscte-iul.pt}           
           \and
           Fernando Brito e Abreu \at
              ISTAR-Iscte, Instituto Universitário de Lisboa, Lisboa, Portugal\\
              \email{fba@iscte-iul.pt}
            \and
            Glauco de Figueiredo Carneiro \at
              Universidade Salvador (UNIFACS), Salvador, Bahia, Brazil\\
              \email{glauco.carneiro@unifacs.br}
}

\date{Received: date / Accepted: date}

\maketitle

\begin{abstract}

Code smells are seen as major source of technical debt and, as such, should be detected and removed. However, researchers argue that the subjectiveness of the code smells detection process is a major hindrance to mitigate the problem of smells-infected code.

We proposed the crowdsmelling approach based on supervised machine learning techniques, where the wisdom of the crowd (of software developers) is used to collectively calibrate code smells detection algorithms, thereby lessening the subjectivity issue. This paper presents the results of a validation experiment for the crowdsmelling approach. 

In the context of three consecutive years of a Software Engineering course, a total ``crowd'' of around a hundred teams, with an average of three members each, classified the presence of 3 code smells (Long Method, God Class, and Feature Envy) in Java source code. These classifications were the basis of the oracles used for training six machine learning algorithms. Over one hundred models were generated and evaluated to determine which machine learning algorithms had the best performance in detecting each of the aforementioned code smells. 

Good performances were obtained for God Class detection (ROC=0.896 for Naive Bayes) and Long Method detection (ROC=0.870 for AdaBoostM1), but much lower for Feature Envy (ROC=0.570 for Random Forrest).

Obtained results suggest that crowdsmelling is a feasible approach for the detection of code smells, but further validation experiments are required to cover more code smells and to increase external validity.

\keywords{crowdsmelling \and code smells \and code smells detection \and software quality \and software maintenance \and collective knowledge \and machine learning algorithms}

\end{abstract}


\section{Introduction}
\label{sec:Introduction}

Maintenance tasks are incremental modifications to a software system that aim to add or adjust some functionality or to correct some design flaws and fix some bugs. It has been found that feature addition, modification, bug fixing, and design improvement can cost as much as 80\% of total software development cost \citep{Travassos1999}. In addition, it is shown that software maintainers spend around 60\% of their time in understanding code \citep{Zitzler2003}. Therefore, as much as almost half (80\%x60\%=48\%) of total development cost may be spent on understanding code. This high cost can be reduced by the availability of tools to increase code understandability, adaptability, and extensibility \citep{Mansoor2017}.

In software development and maintenance, especially in complex systems, the existence of code smells jeopardizes the quality of the software and hinders several operations such as code reuse. Code smells are not bugs, since they do not prevent a program from functioning, but rather symptoms of software maintainability problems \citep{Yamashita2013}. They often correspond to the violation of fundamental design principles and may slow down software evolution (e.g. due to code misunderstanding) or increase the risk of bugs or failures in the future. Code smells can then compromise software quality in the long term by inducing technical debt \citep{Bavota2016}.

Many techniques and tools have been proposed in the literature for detecting code smells \citep{Reis2020}, but that detection faces a few challenges. The first is that code smells lack a formal definition \citep{Wang2015}. Therefore, their detection is highly subjective (e.g. dependent on the developer's experience). Second, due to the dramatic growth in the size and complexity of software systems in the last four decades \citep{Humphrey2009}, it is not feasible to detect code smells thoroughly without tools.

Several approaches and tools for detecting code smells have been proposed. \cite{Kessentini2014} classified those approaches into 7 categories: metric-based approaches, search-based approaches, symptom-based approaches, visualization based approaches, probabilistic approaches, cooperative based approaches, and manual approaches. The most popular code smells detection approach is metric-based. The latter is based on the application of detection rules that compare the values of relevant metrics extracted from the source code with empirically identified thresholds. However, these techniques present some problems, such as subjective interpretation, a low agreement between detectors \citep{Fontana2012}, and threshold dependability.

To overcome the aforementioned limitations of code smell detection, researchers recently applied supervised machine learning techniques that can learn from previous datasets without needing any threshold definition. The main impediment for applying those techniques is the scarcity of publicly available oracles, i.e. tagged datasets for training detection algorithms. To mitigate this hindrance, we have proposed \textit{CrowdSmelling} \citep{Reis2017}, a collaborative crowdsourcing approach, based in machine learning, where the wisdom of the crowd (of software developers) is used to collectively calibrate code smells detection algorithms. The applications based in collective intelligence, where the contribution of several users allows attaining benefits of scale and/or other types of competitive advantage, are gaining increasing importance in Software Engineering \citep{Stol2014} and other areas \citep{Bigham2014,Bentzien2013}. The most notable examples of crowdsourcing in Software Engineering are \textit{crowdtesting} \citep{Sharma2014} and \textit{code snippets recommendation} \citep{Proksch2014}.

In this paper we present the first results of applying \textit{CrowdSmelling}. The paper is organized as follows: next section introduces the related work; then, section \ref{sec:StudyDesign} describes the study design; results and corresponding analyses and the answers to the research questions are presented in section \ref{sec:Results}; threats to the validity of our results are presented in section \ref{sec:Threatsvalidity}; and the concluding remarks, as well as scope for future research, are presented in section \ref{sec:Conclusion}.


\section{Related Work}
\label{sec:RelatedWork}
To the best of our knowledge, there is no study that uses a collective knowledge-based approach to detect code smells automatically, i.e. based on machine leaning, with a dataset increment over 3 years. The use of groups of people in code smells detection is typically used in manual detection approaches and in the construction of oracles (a tagged dataset for training detection algorithms). However, manual techniques are human-centric, tedious, time-consuming, and error-prone. These techniques require a great human effort, therefore not effective for detecting CS in large systems. 

\cite{Tahir2018} presented a study where they investigate how developers discuss code smells and anti-patterns over Stack Overflow to understand better their perceptions and understanding of these two concepts. In this paper, both quantitative and qualitative techniques were applied to analyze discussions containing terms associated with code smells and anti-patterns. The authors reached conclusions like: i) developers widely use Stack Overflow to ask for general assessments of code smells or anti-patterns, instead of asking for particular refactoring solutions, ii) developers very often ask their peers ‘to smell their code’ (i.e., ask whether their own code ‘smells’ or not), and thus, utilize Stack Overflow as an informal, crowd-based code smell/anti-pattern detector, iii) developers often discuss the downsides of implementing specific design patterns, and ‘flag’ them as potential anti-patterns to be avoided. Conversely, we found discussions on why some anti-patterns previously considered harmful should not be flagged as anti-patterns, iv) C\#, JavaScript and Java were the languages with most questions on code smells and anti-patterns, constituting 59\% of the total number of questions on these topics, v) Blob, Duplicated Code and Data Class are the most frequently discussed smells in Stack Overflow, vi) when authors analyzed temporal trends in posts on code smells and anti-patterns in Stack Overflow, show that there has been a steady increase in the numbers of questions asked by developers over time. 

\cite{Palomba2015} presented LANDFILL, a Web-based platform for sharing code smell datasets, and a set of APIs for programmatically accessing LANDFILL’s contents. This platform was created due to the lack of publicly available oracles (sets of annotated code smells). The web-based platform has a dataset of 243 instances of five types of code smells (Divergent Change, Shotgun Surgery, Parallel Inheritance, Blob, and Feature Envy) identified from 20 open source software projects and a systematic procedure for validating code smell datasets. LANDFILL allows anyone to create, share, and improve code smell datasets.

\cite{Oliveira2016} performed a controlled experiment involving 28 novice developers, aimed at assessing the effectiveness of collaborative practices in the identification of code smells. The authors used Pair Programming (PP) and Coding Dojo Randori (CDR), which are two increasingly adopted practices for improving the effectiveness of developers with limited or no knowledge in Software Engineering tasks, including code review tasks, and compared this two practices (PP and CDR) with solo programming in order to better distinguish their impact on the effective identification of code smells. The results suggest that collaborative practices contribute to the effectiveness on the identification of a wide range of code smells. for nearly all types of inter-class smells, the average of smells identified by novice pairs or groups outperformed at least in 40\% of the corresponding average of smells identified by individuals and collaborative practices tend to increase the rate of success in identifying more complex smells. In the same year \cite{Oliveira2016a} performed a research  based on a set of controlled experiments conducted with more than 58 novice and professional developers, with the aim of knowing how to improve the efficiency in the collaborative identification of code smells, and reached the same conclusions as the first study.

\cite{Oliveira2017} is this paper reports an industrial
case study aimed at observing how 13 developers individually and collaboratively performed smell identification in five software projects from two software development organizations. The results are in line with previous studies by these author, where they suggest that collaboration contributes to improving effectiveness on the identification of a wide range of code smells.

\cite{Mello2017} presents and discusses a set of context factors that may influence the effectiveness of smell identification tasks.
The authors presented an initial set of practical suggestions for composing more effective teams to the identification of code smells. These suggestions are, i) be sure all team professionals are aware of the code smell concepts applied in the review, ii) be sure all team professionals are aware of the relevance of identifying code smells, iii) take preference to use collaboration in the reviews, iv) include professionals that had worked in the module and professionals without such experience, v) include professionals with different professionals roles.

\cite{OLIVEIRA2020} have carefully designed and conducted a controlled experiment with 34 developers. The authors exploited a particular scenario that reflects various organizations: novices and professionals inspecting systems they are unfamiliar with. They expect to minimize some critical threats to validity of previous work. Additionally, they interviewed 5 project leaders aimed to understand the potential adoption of the collaborative smell identification in practice. Statistical testing suggests 27\% more precision and 36\% more recall through the collaborative smell identification for both novices and professionals. The interviews performed by the authors showed that leaders would strongly adopt the collaborative smell identification. However, some organization and tool constraints may limit such adoption.

\cite{Baltes2020} presented a study with similarities  and differences between code clones in general and code clones on Stack Overflow and point to open questions that need to be addressed to be able to make data-informed decisions about how to properly handle clones on this important platform. The results of his first preliminary investigation indicated that clones in Stack Overflow are common, diverse, similar to clones in regular software projects, affect the maintainability of posts and can lead to licensing issues. The authors further point to specific challenges, including incentives for users to clone successful answers and difficulties with bulk edits on the platform.

Regarding the use of the machine learning approach in the detection of code smells, most studies only use one algorithm, being the most usual algorithm the decision trees. We will present below the most relevant studies that use multiple machine learning algorithms.

Some of the most relevant studies in the area of machine learning were performed by \cite{Fontana2013,Fontana2015}. In the first work \cite{Fontana2013}, they outlined some common problems of code smell detectors and described the approach they were following based on machine learning technology. In this study the authors focused on 4 code smells (Data Class, Large Class, Feature Envy, Long Method), considered 76 systems for analysis and validation and experimented 6 different machine learning algorithms. The results with a use 10-fold cross-validation to assess the performance of predictive models shown that J48, Random Forest, JRip and SMO have accuracy values greater than 90\% for the 4 code smells, and on average they have the best performances. In the second work \cite{Fontana2015}, they performed the largest experiment of applying machine learning algorithms. They experimented 16 different machine-learning algorithms on four code smells (Data Class, Large Class, Feature Envy, Long Method) and 74 software systems, with 1986 manually validated code smell samples. They found that all algorithms achieved high performances in the cross-validation data set, yet the highest performances were obtained by J48 and Random Forest, while the worst performance were achieved by support vector machines. The authors concluded that the application of machine learning to the detection of these code smells can provide high accuracy (\textgreater96 \%), and only a hundred training examples are needed to reach at least 95 \% accuracy. The authors interpret the results as an indication that “using machine learning algorithms for code smell detection is an appropriate approach”.

\cite{Nucci2018} replicated the \cite{Fontana2015} study  with a different dataset configuration. The dataset contains instances of more than one type of smell, with a reduced proportion of smelly components and with a smoothed boundary between the metric distribution of smelly and non-smelly components, and therefore more realistic. The results revealed that with this configuration the machine learning techniques reveal critical limitations in the state of the art which deserve further research. They concluded that, when testing code smell prediction models on the revised dataset, they noticed: i) accuracy of all the models is still noticeable high when compared to the results of the reference study (on average, 76\% vs 96\%), ii) that performances are up to 90\% less accurate in terms of F-Measure than those reported in the Fontana et al. study. Thus, the problem of detecting code smells through the adoption of machine learning techniques may still be worthy of further attention, e.g., in devising proper machine learning-based code smell detectors and datasets for software practitioners.


\section{Study design}
\label{sec:StudyDesign}
The concept of crowdsmelling – use of collective intelligence in the detection of code smells – aims to mitigate the aforesaid problems of subjectivity and lack of calibration data required to obtain accurate detection model parameters. Thus, our approach consists of several teams using a tool to detect code smells and then confirming the validity of the detection manually. In addition to the code smells detected by the tool, teams can always add other code smells manually. In the end, code identification, code metrics and classification (presence or absence of code smells) are saved by creating an oracle for each code smell. This oracle will allow training algorithms for code smells detection. These oracles have been increased for 3 years, with data collected each year.

The repetition of this process for 3 years, allowed every year to increase the oracle with data from new teams, thus increasing the variability of existing classifications. This variability of opinions in the code smells classification is very important, because it will allow collecting data from teams with different opinions on the definition of code smells, enriching the oracle.  

\subsection{Subject Selection}
\label{Subject}
Our subjects were the finalists (3rd year) of a B.Sc. degree on computer science at the Iscte-IUL university, attending a compulsory Software Engineering course. They had similar backgrounds as they have been trained
across the same set of courses along their academic path. The knowledge about code smell was acquired in the Software Engineering curricular unit.

\begin{table} [h]
\centering
\caption{Teams that did the detection of code smells }
\label{table:TeamsDetectionCS}
\begin{tabular}{ccc}
\hline\noalign{\smallskip}
Year & Number of teams & Total number of elements\\
\noalign{\smallskip}\hline\noalign{\smallskip}
2018 & 8 & 31 \\
2019 & 51 & 152 \\
2020 & 44 & 179 \\
\noalign{\smallskip}\hline
\end{tabular}
\end{table}

The teams had a variable size depending on the year (see Table \ref{table:TeamsDetectionCS}) and the number of participants were increasing each year. In 2018, 8 teams were formed, essentially with 4 members each, for a total of 31 elements. In 2019 we had 51 teams, mainly made up of 3 members, with a total of 152 members. In 2020 we had 44 teams, mainly made up of 6 members, with a total of 179 members. These teams were requested to complete a code smells detection assignment.

\subsection{Data}
\label{Data}
The participants were invited to perform the detection of 3 code smells (God Class, Feature Envy, Long Method), having JDeodorant\footnote{https://users.encs.concordia.ca/~nikolaos/jdeodorant/} as an auxiliary tool in the detection, but each team is free to choose the code it wants to classify and add code smells manually regardless of the tool. The applications used on code smells detection were jasml-0.10\footnote{http://jasml.sourceforge.net/}, jgrapht-0.8.1\footnote{https://jgrapht.org/} and jfreechart-1.0.13\footnote{https://www.jfree.org/} in 2018, in the next 2 years it was just Jasml for a better experience control.

The results of each team's detection is saved in a file with the following fields for the code smells Feature Envy and Long Method: Team number, project, package, class, method, 82 metrics of code, is code smell. In the case of the code smell God Class, as the scope is to the class, does not have the method field and 61 code metrics are saved. At the end we have 3 files, one for each smell code. 

The data obtained each year serve to reinforce the calibration datasets of the machine learning algorithms, with the objective of improving their detection performance over time. This way we will have several datasets, so we can evaluate which one gives the best results for each code smell.

\begin{table}[h]
\centering
\caption{Datasets (Oracles) and their composition}
\label{table:Datasets}
\begin{tabular}{lcccccc}
\hline\noalign{\smallskip}
\multicolumn{1}{c}{Dataset} & Code smell   & Nº Cases & True & \% True & False & \% False \\
\noalign{\smallskip}\hline\noalign{\smallskip}
2018                        & Feature Envy & 10           & 3    & 30\%    & 7     & 70\%     \\
2019                        & Feature Envy & 197          & 110  & 56\%    & 87    & 44\%     \\
2019+2018                   & Feature Envy & 207          & 113  & 55\%    & 94    & 45\%     \\
2020                        & Feature Envy & 123          & 79   & 64\%    & 44    & 36\%     \\
2020+2019                   & Feature Envy & 320          & 189  & 59\%    & 131   & 41\%     \\
2020+2019+2018              & Feature Envy & 330          & 192  & 58\%    & 138   & 42\%     \\
2018                        & God class    & 22           & 8    & 36\%    & 14    & 64\%     \\
2019                        & God class    & 129          & 74   & 57\%    & 55    & 43\%     \\
2019+2018                   & God class    & 151          & 82   & 54\%    & 69    & 46\%     \\
2020                        & God class    & 136          & 84   & 62\%    & 52    & 38\%     \\
2020+2019                   & God class    & 265          & 158  & 60\%    & 107   & 40\%     \\
2020+2019+2018              & God class    & 287          & 166  & 58\%    & 121   & 42\%     \\
2018                        & Long Method  & 59           & 24   & 41\%    & 35    & 59\%     \\
2019                        & Long Method  & 414          & 180  & 43\%    & 234   & 57\%     \\
2019+2018                   & Long Method  & 473          & 204  & 43\%    & 269   & 57\%     \\
2020                        & Long Method  & 853          & 350  & 41\%    & 503   & 59\%     \\
2020+2019                   & Long Method  & 1267         & 530  & 42\%    & 737   & 58\%     \\
2020+2019+2018              & Long Method  & 1326         & 554  & 42\%    & 772   & 58\%     \\
\noalign{\smallskip}\hline
\end{tabular}
\end{table}

In Table \ref{table:Datasets} we present the composition of the datsets, indicating the following elements, i) name of the dataset, ii) code smell to which the dataset refers, iii) number of cases, iv) number of true instances, v) percentage of true instances, vi) number of false instances, vii) percentage of false instances. Each dataset is identified by the year or the years that constitute it, for example, 2019 is the dataset of the year 2019 and 2019+2020 is the dataset resulting from the aggregation of the datasets of the years 2019 and 2020. Unlike several authors, such as \cite{Fontana2015}, we do not normalize our datasets in size in order to balance the number of positive and negative instances. Even with the risk of getting worse results, we used the datasets with all the cases classified by the teams. Thus, we believe that we are reproducing the reality of the teams' thinking about code smells. 
The size of the datasets varies widely depending on the type of code smell. 
Since the datasets of code smell Feature Envy are too small it was not possible to obtain good results. For the dataset 2018 of Feature Envy it was not possible to obtain precision, and consequently f-measure, since all the instances classified as TRUE were poorly classified, i.e., all the instances were classified as FALSE. For the dataset 2020, we also did not obtain precision and f-measure, because all the instances classified as FALSE were badly classified, i. e., all the models created from this dataset to classify the future envy, classified all the instances of the dataset 2020 as TRUE. Even so, we intend to use all the datasets, as they represent the obtained reality and serve as a basis for a future amplification and evolution of the crowd's study in code smells detection.

\subsection{Code Smells}
\label{CodeSmells}
In this study we considered three different types of code smells defined by \cite{Fowler1999}:
\begin{itemize}
  \item God Class.This smell characterises classes having a large size, poor cohesion, and several dependencies with other data classes of the system. Class that has many responsibilities and therefore contains many variables and methods. The same Single Responsibility Principle (SRP) also applies in this case;
  \item Feature Envy. When a method is more interested in members of other classes than its own, is a clear sign that it is in the wrong class;
  \item Long Method. Methods implementing more than one functionality are affected by this smell. Very large method/function and, therefore, difficult to understand, extend and modify. It is very likely that this method has too many responsibilities, hurting one of the principles of a good Object Oriented design (SRP: Single Responsibility Principle);
\end{itemize}

The choice of these 3 code smells is due to the fact that, according to the Systematic Literature Review we conducted, they are the three most detected code smells. Therefore, it is easier for teams to obtain documentation and understand these 3 code smells for better detection.

\subsection{Code Metrics}
\label{CodeMetrics}
In this study, we used the same metrics that were used in the study of \cite{Fontana2015}, since the metrics are publicly available.

The metrics extracted from the software which constitute the independent variables in the machine learning algorithms, are at class, method, package and project level. For God Class, we used a set of 61 metrics, and for the other two code smells, Feature Envy and Long Method, we used a set of 82 metrics, plus 21 metrics than God Class, since these codes smells are at the method level. The main metrics are described in the table \ref{AppendixA}.

\subsection{Machine Learning Techniques Experimented}
\label{MLTechniques}
The application used in this experiment to train and evaluate machine learning algorithms was Weka (open source software from Waikato University) \citep{Hall2009}, and the following algorithms available in Weka were implemented:

\begin{itemize}
\item J48 \citep{Quinlan2014} is an implementation of the C4.5 decision tree, and its three types of pruning techniques: pruned, unpruned and reduced error pruning;
\item Random Forest \citep{Breiman2001} consists of a large number of individual decision trees, a forest of random trees, that operate as an ensemble;
\item AdaBoostM1 \citep{Freund1996} Boosting works by repeatedly running a given weak learning algorithm on various distributions over the training data, and then combining the classifiers produced by the weak learner into a single composite classifier. Weka uses the Adaboost M1 method;
\item SMO \citep{Platt1999} is a Sequential Minimal Optimization algorithm widely used for training support vector machines. We use the Polynomial kernel;
\item Multilayer Perceptron \citep{Rumelhart1986} is a classifier that uses backpropagation to learn a multi-layer perceptron to classify instances;
\item Naïve Bayes \citep{George1995} is a probabilistic model based on the Bayes theorem.
\end{itemize}

Experiments were performed to evaluate the performance values of the machine learning algorithms used with their default parameters for each type of code smell.

\subsection{Model Evaluation}
\label{Evaluation}
To assess the capabilities of the machine learning model, we adopted 10-Fold Cross Validation \citep{Stone1974}. This methodology randomly partitions the data into 10 folds of equal size, applying a stratified sampling (e.g., each fold has the same proportion of code smell instances). A single fold is used as test set, while the remaining ones are used as training set. The process was repeated 10 times,using each time a different fold as test set. The result of the process described above consisted of a confusion matrix for each code smell type and for each model \citep{Pecorelli2019}.

Several evaluation metrics can be used to assess model quality in terms of false positives/negatives (FP/FN), and true classifications (TP/TN). However, commonly used measures, such as \textbf{Accuracy, Precision, Recall and F-Measure}, do not perform very well in case of an imbalanced dataset or they require the use of a minimum probability threshold to provide a definitive answer for predictions. For these reasons, we used the \textbf{ROC}\footnote{Receiver operating characteristic (\textbf{ROC}) is a curve that plots the true positive rates against the false positive rates for all possible thresholds between 0 and 1.}, which is a threshold invariant measurement. Nevertheless, for general convenience, we kept present in results tables all the evaluation metrics \citep{caldeira2020}.

\begin{equation}
Accuracy = \frac{TP + TN}{TP + FP + FN + TN}
\end{equation}
\begin{equation}
Precision = \frac{TP}{TP + FP}
\end{equation}
\begin{equation}
Recall = \frac{TP}{TP + FN}
\end{equation}
\begin{equation}
F-measure =  2*\frac{Recall * Precision}{Recall + Precision}
\end{equation}

\textbf{ROC} gives us a 2-D curve, which passes through (0, 0) and (1, 1). The best possible model would have the curve close to y = 1, with and area under the curve (\textbf{AUC}) close to 1.0. \textbf{AUC} always yields an area of 0.5 under random-guessing. This enables comparing a given model against random prediction, without worrying about arbitrary thresholds, or the proportion of subjects on each class to predict \citep{Rahman2013}.

\subsection{Process}
\label{Process}
For a better understanding of this exploratory study, we will describe the 3 stages that constitute the process.

\subsubsection{Stage 1: Developer - Code smell classification}
\label{subsub:Stage1}
All Java developers use the Eclipse IDE with the JDeodorant plug-in installed. In the first year - the year 2018 - each team was free to choose, from a list of java projects, the one they wanted to use to detect Code smell.  So, in the first year, the teams chose the java projects jasml-0.10, jgrapht-0.8.1, and jfreechart-1.0.13. In the following two years, the teams just used jasml-0.10. 

\begin{figure}
\centering
\includegraphics [trim= 1in .5in 1in 1.2in, clip,scale=0.5] {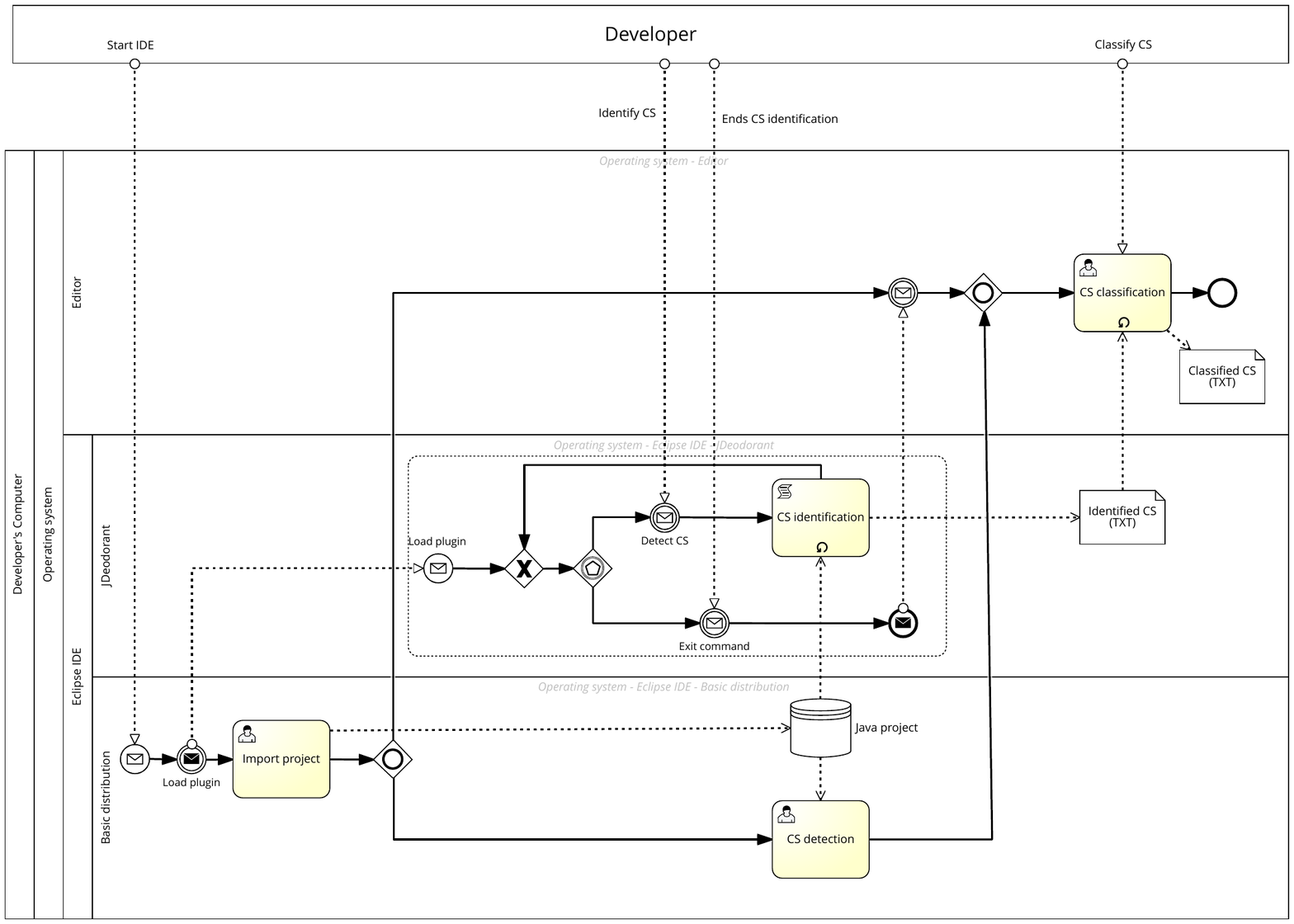}
\caption{Process of code smells classification by the developer}
\label{fig:ProcessDeveloper}
\end{figure}

Figure \ref{fig:ProcessDeveloper} shows the cod smells classification process by the programmer, where we can see that after importing the java project, the participants were invited to perform the detection of 3 Code smell (Long Method, God Class, Feature Envy).After importing the java projects, the participants were invited to perform the detection of 3 code smells (GodClass, Feature Envy, Long Method). In this detection the participants could use JDeodorant as an auxiliary tool in the detection of smells, i.e., they first used JDeodorant as an advisor, and then manually validated the result of the detection of the tool, saying whether or not they agreed with the code smells detected. The use of JDeodorant also had the advantage that participants could export the code smells identified by this tool to a text file, where they later registered their agreement or not with this identification, i.e., they performed the final classification. 

Regardless of the use of JDeodorant, all participants could identify the code smells directly in the java project code and record their occurrence or not in a text file. In this case, the participants wrote in the text file the name of the class or method, and if there existed or not a code smell. 

As a result of this stage, all teams produced three files - one for each code smells - with the classification of a set of methods and classes of the java project, i.e., with the record of the existence or not of code smells in those classes or methods. This stage was performed over 3 years, 2018, 2019 and 2020.

\subsubsection{Stage 2: Researcher - Evaluation of machine learning models}
\label{subsub:Stage2}
\begin{figure}
\centering
\includegraphics [trim= 1in 2.5in 1in 1.2in, clip, scale=0.4] {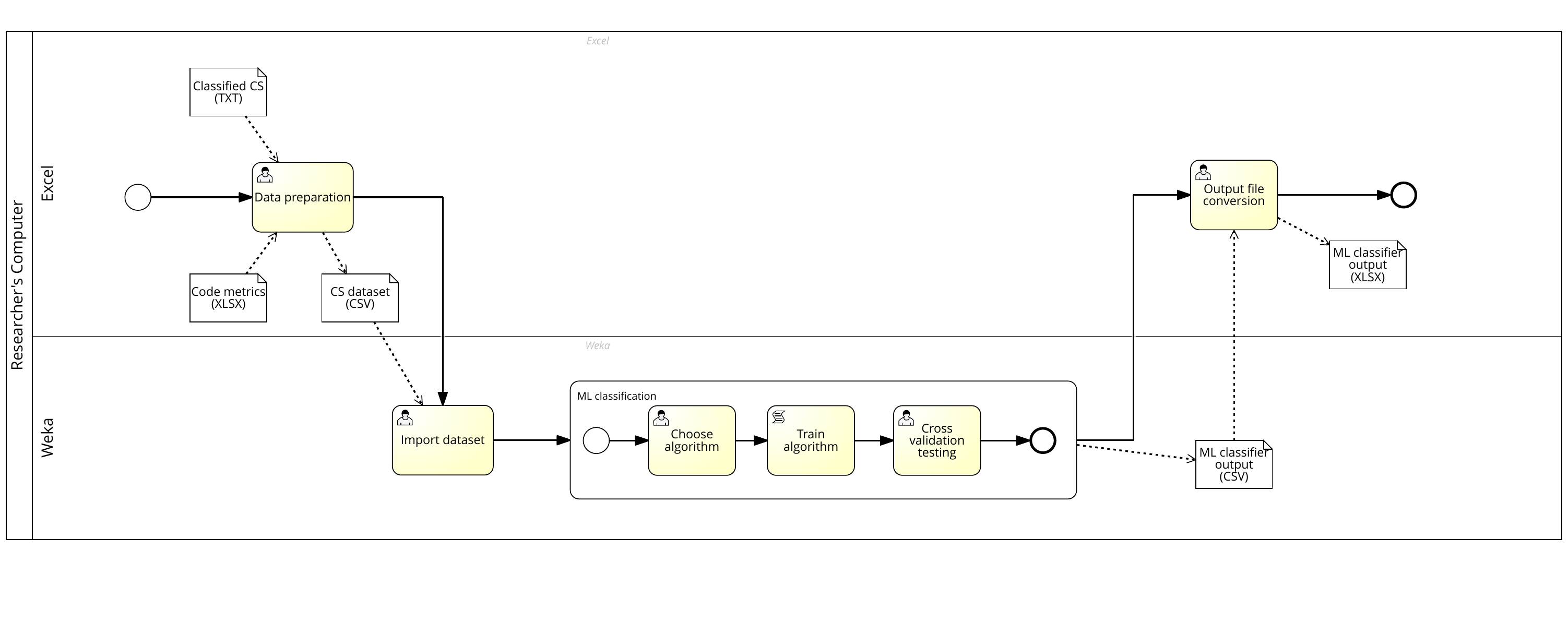}
\caption{Process of creation of the datasets and evaluation of the machine learning techniques by the researcher}
\label{fig:Processresearcher1}
\end{figure}

After collecting data in three years, we proceeded to the second phase, which aimed to produce the datasets for the 3 code smells and evaluation of the different machine learning techniques. In figure \ref{fig:Processresearcher1} is represented the whole process of this second stage.

The first task to be performed by the researcher is the creation of the datasets described in section \ref{Data}. 

The creation of the datasets is done by joining all the text files with the classifications of a code smell, produced by the teams of each year, in a single Excel file. Then, to this excel file are added the code metrics for the methods or classes (see section \ref{CodeMetrics}), depending on the scope of the code smell to which the dataset belongs. Thus, in a first step, datasets are created - usually called oracles - with the data for each year, for each of the three code smells, for a total of 6 datasets. These datasets have been given the name of the year to which they belong, i.e., 2018, 2019 and 2020. In a second step, we proceed to aggregate the dataset of the year with those of previous years to make the dataset larger, increasing the number of instances. In the end, we created six datasets for each code smell, with a total of 36 datasets (see table \ref{table:Datasets}).

After creating the datasets, we proceed to the creation and evaluation of the machine learning models using Weka (open source software from Waikato University) \citep{Hall2009}. To import datasets into Weka, we convert the datasets files, from excel XLSX to CSV. At Weka, we trained the six algorithms described in section \ref{MLTechniques}, with each of the 36 datasets, and evaluated the model produced using the 10-Fold Cross Validation methodology. In the end, 36 machine learning models were created for each code smell, with a total of 108 models for the three code smells. Finally, all the metrics (Accuracy, Precision, Recall, F-Measure, and ROC) resulting from the evaluation of each model were saved in the "ML classifier output" file (see section \ref{Evaluation}).

\subsubsection{Stage 3: Researcher - Model variance test}
\label{subsub:Stage3}
\begin{figure}
\centering
\includegraphics [trim= 4in 3in 4in 1.3in, clip, scale=0.5] {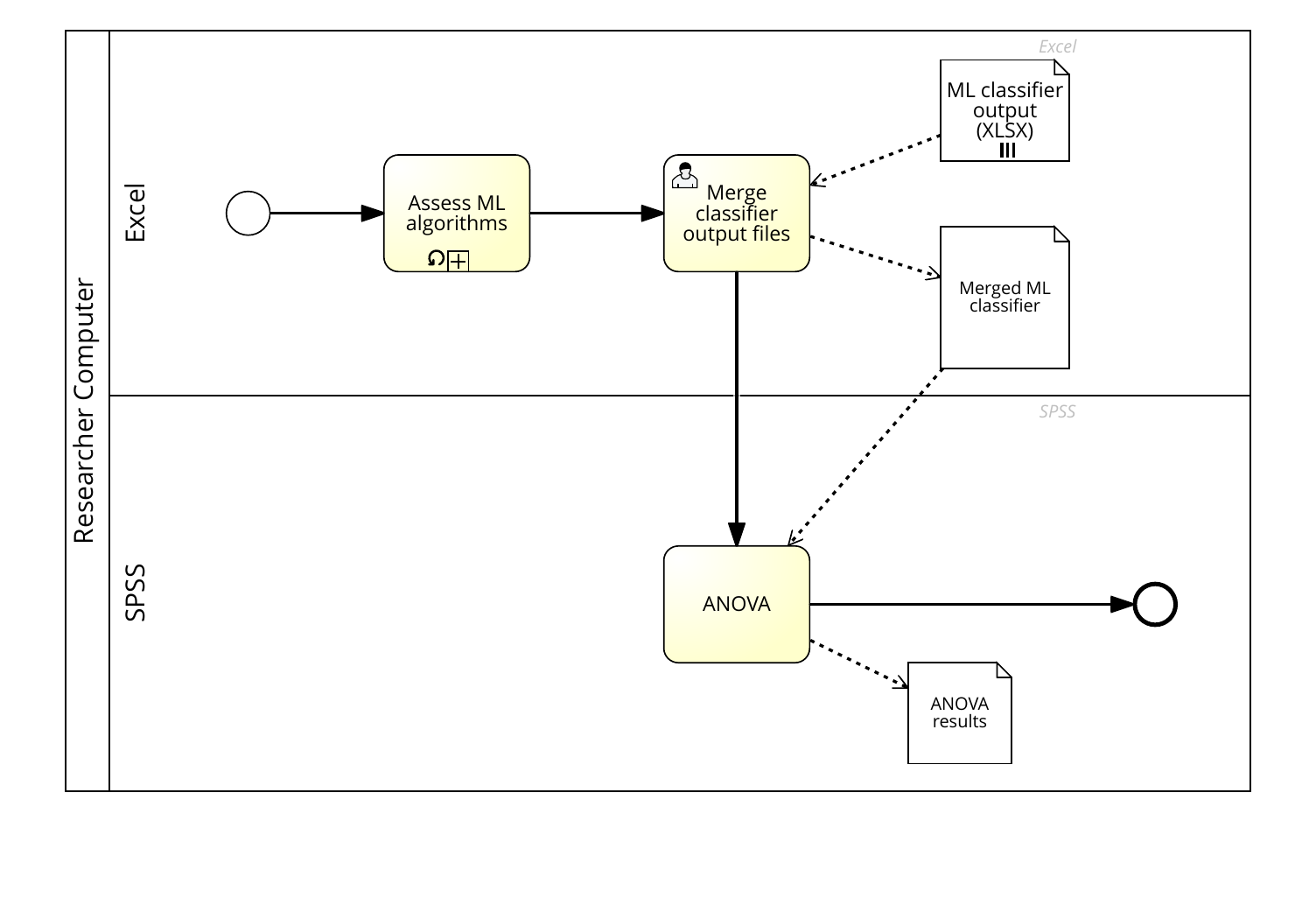}
\caption{Process of testing the variance between machine learning models}
\label{fig:fig:Processresearcher2}
\end{figure}

To check if there were significant differences between the classifications presented by the different models, we proceeded to the analysis of variance through a one-way analysis of variance (ANOVA) (see figure \ref{fig:fig:Processresearcher2}).

To test the variance between the machine learning models we use the ROC value. Thus, the first step was to produce a data file, for each code smell, with the identification of the machine learning models and the respective ROC. This file was created aggregating the results of the evaluations of all models produced by Weka, by code smell. 

To analyze if there were differences between the classifications of the machine learning models for each code smell, we performed an analysis of variance using a one-way analysis of variance (ANOVA) test in the IBM SPSS Statistics software.

\subsection{Research Questions}
\label{ResearchQuestions}
To understand the use of Crowdsmelling - the use of Collective Knowledge in code smells detection - we have formulated the following research questions:
\begin{itemize}

\item[$\bullet$] \textbf{RQ1:} What is the performance of machine learning techniques when trained with data from the crowd and therefore more realistic?   
\item[$\bullet$] \textbf{RQ2:} What is the best machine learning model to detect each one of the three code smells?
\item[$\bullet$] \textbf{RQ3:} It is possible to use Collective Knowledge for code smells detection?
\end{itemize}


\section{Results}
\label{sec:Results}
In this section, we present the experiment results with respect to our
research questions.

\subsection{\textbf{RQ1. What is the performance of machine learning techniques when trained with data from the crowd and therefore more realistic? }}
\label{ResultsRQ1}

In this \textbf{RQ} we will evaluate the performance of the 36 models for each code smell in a total of 108 models. These models resulted from the training of the 6 machine learning algorithms (J48, Random Forest, AdaBoostM1, SMO, Multilayer  Perceptron, Naïve Bayes), described in section \ref{MLTechniques}, by the datasets presented in table \ref{table:Datasets}. These algorithms were trained with the various datasets resulting from the crowd, and as explained in \ref{Data} we want these datasets to be as real as possible, to represent as faithfully as possible what the detection teams think about the code smells. 

From the various existing metrics for evaluating machine learning models, we have chosen to use the ROC as the primary metric, but we also use accuracy, precision, recall, and f-measure. For testing, we used the 10-Fold Cross Validation, for the reasons presented in \ref{Evaluation}. 

\subsubsection{Performance of machine learning techniques for code smell Long Method}
\label{ResultsRQ1-LM}
Starting by analyzing the machine learning techniques for the Long Method data, described in Table \ref{table:ROC-LM}, we observed that the best results were obtained by the Random Forrest and AdaBoostM1 algorithms. The best result with a ROC of 0.870 was obtained by AdaBoostM1 when trained by the dataset 2020, followed by the Random Forrest with ROC of 0.869 for the same dataset. For the dataset 2018, the best result was also that of AdaBoostM1. However, the most uniform algorithm was Random Forrest, with the best results in 4 of the 6 datasets (2020+2019+2018, 2020+2019, 2019+2018, 2019) and for the dataset 2020, the difference for AdaBoostM1 is insignificant (0.001).
The Multilayer Perceptron and J48 algorithms, were two other algorithms to present ROC results above 0.800. Especially the Multilayer Perceptron algorithm which for the datasets of the year 2020 presented an ROC between 0.868 and 0.822. 

\begin{table}[ht]
\centering
\caption{Long Method: ROC Area results for the machine learning algorithms trained by the 3 years datasets}
\label{table:ROC-LM}
\begin{adjustbox}{width=\textwidth}
\begin{tabular}{lcccccc}
\hline\noalign{\smallskip}
\multicolumn{1}{r}{year} & \multicolumn{3}{c}{2020} & \multicolumn{2}{c}{2019} & \multicolumn{1}{c}{2018} \\
\cmidrule(lr){2-4} \cmidrule(lr){5-6} \cmidrule(lr){7-7}
\diagbox{Algorithm}{dataset} &  2020+2019+2018 & 2020+2019 & 2020 & 2019+2018 & 2019 & 2018 \\ 
 \noalign{\smallskip}\hline\noalign{\smallskip}
J48 & 0.792 & 0.801 & 0.832 & 0.677 & 0.678 & 0.617 \\
Random Forest & 0.828 & 0.828 & 0.869 & 0.684 & 0.679 & 0.671 \\
AdaBoostM1 & 0.807 & 0.818 & \cellcolor[HTML]{EFEFEF}0.870 & 0.665 & 0.673 & 0.707 \\
SMO & 0.753 & 0.753 & 0.803 & 0.634 & 0.649 & 0.524 \\
MultilayerPerceptron & 0.822 & 0.822 & 0.868 & 0.683 & 0.667 & 0.604 \\
NaiveBayes & 0.736 & 0.742 & 0.783 & 0.584 & 0.614 & 0.471 \\
\noalign{\smallskip}\hline
\end{tabular}
\end{adjustbox}
\end{table}

The worst results were obtained by the NaiveBayes algorithm with ROC between 0.783 and 0.471. The second worst algorithm was SMO, with ROC results between 0.803 and 0.524.

In table \ref{table:ROC-LM}, we can still observe that the best results were obtained when the algorithms were trained with the datasets for the year 2020, with ROC of 0.870 for the dataset 2020 and ROC of 0.828 for the datasets 2020+2019+2018 and 2020+2019. In opposition is the year 2019, with the worst results, ROC of 0.684 and 0.679 for the datasets 2019+2018, 2019, respectively.

\subsubsection{Performance of machine learning techniques for code smell God Class}
\label{ResultsRQ1-GC}
Table \ref{table:ROC-GC} shows the results of the machine learning techniques for the God Class data. The best result was obtained by the NaiveBayes algorithm, when trained by the dataset 2020, with the ROC value of 0.896. The algorithms that obtained the best performances were NaiveBayes and MultilayerPerceptron, with the best result in 3 of the datasets each one. NaiveBayes obtained the best results for the datasets 2020, 2020+2019, 2019, with ROC values of 0.896, 0.859 and 0.804, respectively.Also with the best result in 3 datasets (2020+2019+2018, 2019+2018, 2018) the MultilayerPerceptron algorithm presented ROC values between 0.768 and 0.885. The Random Forest and AdaBoostM1 algorithms presented their best ROC values of 0.893 and 0.876, respectively, for the dataset 2020.

\begin{table}[ht]
\centering
\caption{God Class: ROC Area results for the machine learning algorithms trained by the 3 years datasets}
\label{table:ROC-GC}
\begin{adjustbox}{width=\textwidth}
\begin{tabular}{lcccccc}
\hline\noalign{\smallskip}
\multicolumn{1}{r}{year} & \multicolumn{3}{c}{2020} & \multicolumn{2}{c}{2019} & \multicolumn{1}{c}{2018} \\
\cmidrule(lr){2-4} \cmidrule(lr){5-6} \cmidrule(lr){7-7}
\diagbox{Algorithm}{dataset} &  2020+2019+2018 & 2020+2019 & 2020 & 2019+2018 & 2019 & 2018 \\ 
 \noalign{\smallskip}\hline\noalign{\smallskip}
J48 & 0.763 & 0.759 & 0.791 & 0.693 & 0.725 & 0.692 \\
Random Forest & 0.853 & 0.850 & 0.893 & 0.781 & 0.802 & 0.491 \\
AdaBoostM1 & 0.854 & 0.857 & 0.876 & 0.771 & 0.793 & 0.571 \\
SMO & 0.815 & 0.800 & 0.857 & 0.716 & 0.751 & 0.741 \\
MultilayerPerceptron & 0.880 & 0.853 & 0.885 & 0.805 & 0.797 & 0.768 \\
NaiveBayes & 0.731 & 0.859 & \cellcolor[HTML]{EFEFEF}0.896 & 0.669 & 0.804 & 0.651 \\
\noalign{\smallskip}\hline
\end{tabular}
\end{adjustbox}
\end{table}

The worst results were presented by J48 and SMO, with their best ROC values for the dataset 2020 of 0.759 and 0.857, respectively.

Regarding the datasets that presented the best results were those of the year 2020, with the dataset only with data of the year 2020 being the best (dataset 2020) with ROC values between 0.896 and 0.791. The dataset with the worst results was 2018, with the ROC between 0.491 and .0768.

\subsubsection{Performance of machine learning techniques for code smell Feature Envy}
\label{ResultsRQ1-FE}
The ROC Results for the machine learning algorithms trained by the 3-year datasets for the code smell Feature Envy are presented in table \ref{table:ROC-FE}. Feature Envy detection results are low, with the Random Forest algorithm having the best ROC value of 0.570 when trained by dataset 2019. As already explained in point \ref{Data}, the datasets for Feature Envy are very small, however we are convinced that when we have bigger datasets the results will be better. The machine learning algorithms showed better results when trained with the datasets of the year 2019, with ROC values between 0.570 and 0.508.

\begin{table}[ht]
\centering
\caption{Feature Envy: ROC Area results for the machine learning algorithms trained by the 3 years datasets}
\label{table:ROC-FE}
\begin{adjustbox}{width=\textwidth}
\begin{tabular}{lcccccc}
\hline\noalign{\smallskip}
\multicolumn{1}{r}{year} & \multicolumn{3}{c}{2020} & \multicolumn{2}{c}{2019} & \multicolumn{1}{c}{2018} \\
\cmidrule(lr){2-4} \cmidrule(lr){5-6} \cmidrule(lr){7-7}
\diagbox{Algorithm}{dataset} &  2020+2019+2018 & 2020+2019 & 2020 & 2019+2018 & 2019 & 2018 \\ 
 \noalign{\smallskip}\hline\noalign{\smallskip}
J48 & 0.518 & 0.484 & 0.467 & 0.552 & 0.563 & 0 \\
Random Forest & 0.539 & 0.494 & 0.486 & 0.542 & \cellcolor[HTML]{EFEFEF}0.570 & 0 \\
AdaBoostM1 & 0.498 & 0.437 & 0.468 & 0.554 & 0.548 & 0 \\
SMO & 0.520 & 0.491 & 0.500 & 0.551 & 0.508 & 0 \\
MultilayerPerceptron & 0.533 & 0.498 & 0.536 & 0.548 & 0.544 & 0 \\
NaiveBayes & 0.524 & 0.519 & 0.482 & 0.548 & 0.547 & 0 \\
\noalign{\smallskip}\hline
\end{tabular}
\end{adjustbox}
\end{table}

\subsubsection{The one-way analysis of variance (ANOVA)}
\label{ResultsRQ1-ANOVA}
 To determine if there were significant differences between the performance of machine learning techniques when trained with data from the crowd and therefore more realistic, a One-way ANOVA was conducted to compare effect of machine learning techniques on the ROC.  The results obtained were the following:

i) For the code smell Long Method, an analysis of variance showed that the effect of the performance of machine learning techniques on ROC value was not significant, F(5,30)=1.096, p=.383. 

ii) For the code smell God Class, an analysis of variance showed that the effect of the performance of machine learning techniques on ROC value was not significant, F(5,30)=.655, p=.660.

ii) For the code smell Feature Envy, an analysis of variance showed that the effect of the performance of machine learning techniques on ROC value was not significant, F(5,24)=.585, p=.712.

The results of the variance tests show there was no statistically significant difference between the performance of the six machine learning models, when trained with data from the crowd and therefore more realistic.

\subsection{\textbf{RQ2. What is the best machine learning model to detect each one of the three code smells?}}
\label{ResultsRQ2}

In this RQ we want to know which is the best model to detect each of the code smells. To do so, we analyzed the various metrics that evaluate the performance of code smells prediction models in detecting each of the 3 code smells. Of course, the best model will vary with the metric we choose to analyze the model performance (accuracy, precision, recall, f-measure, ROC), but for the reasons described in \ref{Evaluation} we will use as the main metric the ROC.

Tables \ref{table:ResultsLM}, \ref{table:ResultsGC}, and \ref{table:ResultsFE} present the performance of the prediction models for the 3 code smells, where the best values for each of the evaluation metrics are marked.

\subsubsection{Best machine learning model for code smell Long Method}
\label{ResultsRQ2-LM}
For the code smell Long Method, the model that best performs its detection is AdaBoostM1, presenting the best values for all evaluation metrics. As we can see in the table \ref{table:ResultsLM}, AdaBoostM1 obtained a ROC value of 0.870, a accuracy of 81.36\%, a precision of 82.90\%, a recall of 81.40\%, and F-measure of 81.50\%. However, two more models present an almost equal ROC, Random Forest and Multilayer Perceptron, with ROC values of 0.869 and 0.868, respectively. 

Except for NaiveBayes, all the other five models have values higher than 0.803 for ROC and values higher than 80.00\% for f-measure, precision, and recall in the detection of code smell Long Method.

\begin{table}[ht]
\centering
\caption{Long Method: Performance of the code smell prediction models}
\label{table:ResultsLM}
\setlength{\tabcolsep}{4pt}
\begin{adjustbox}{width=\textwidth}
\begin{tabular}{clccccccc}
\hline\noalign{\smallskip}
\textbf{Dataset} & \multicolumn{1}{c}{\textbf{Classifier}} & \textbf{Accuracy} & \textbf{TP Rate} & \textbf{FP Rate} & \textbf{Precision} & \textbf{Recall} & \textbf{F-Measure} & \textbf{ROC Area} \\
\noalign{\smallskip}\hline\noalign{\smallskip}
2018           & J48                  & 61.02\% & 61.00\% & 37.20\% & 63.40\% & 61.00\% & 61.30\% & 0.617 \\
2018           & Random Forest        & 61.02\% & 61.00\% & 45.10\% & 60.00\% & 61.00\% & 60.00\% & 0.671 \\
2018           & AdaBoostM1           & 67.80\% & 67.80\% & 36.50\% & 67.30\% & 67.80\% & 67.40\% & 0.707 \\
2018           & SMO                  & 55.93\% & 55.90\% & 51.20\% & 54.30\% & 55.90\% & 54.50\% & 0.524 \\
2018           & MultilayerPerceptron & 57.63\% & 57.60\% & 46.10\% & 57.40\% & 57.60\% & 57.50\% & 0.604 \\
2018           & NaiveBayes           & 61.02\% & 61.00\% & 47.70\% & 59.40\% & 61.00\% & 58.70\% & 0.471 \\
2019           & J48                  & 64.73\% & 64.70\% & 34.10\% & 66.10\% & 64.70\% & 64.90\% & 0.678 \\
2019           & Random Forest        & 66.18\% & 66.20\% & 34.50\% & 66.40\% & 66.20\% & 66.30\% & 0.679 \\
2019           & AdaBoostM1           & 66.67\% & 66.70\% & 29.60\% & 70.70\% & 66.70\% & 66.30\% & 0.673 \\
2019           & SMO                  & 65.46\% & 65.50\% & 35.70\% & 65.50\% & 65.50\% & 65.50\% & 0.649 \\
2019           & MultilayerPerceptron & 63.29\% & 63.30\% & 38.60\% & 63.10\% & 63.30\% & 63.10\% & 0.667 \\
2019           & NaiveBayes           & 61.11\% & 61.10\% & 40.80\% & 60.90\% & 61.10\% & 61.00\% & 0.614 \\
2019+2018      & J48                  & 65.75\% & 65.80\% & 33.20\% & 67.00\% & 65.80\% & 65.90\% & 0.677 \\
2019+2018      & Random Forest        & 65.33\% & 65.30\% & 35.30\% & 65.60\% & 65.30\% & 65.40\% & 0.684 \\
2019+2018      & AdaBoostM1           & 66.60\% & 66.60\% & 29.10\% & 71.40\% & 66.60\% & 66.20\% & 0.665 \\
2019+2018      & SMO                  & 63.85\% & 63.80\% & 37.10\% & 64.00\% & 63.80\% & 63.90\% & 0.634 \\
2019+2018      & MultilayerPerceptron & 63.00\% & 63.00\% & 39.40\% & 62.70\% & 63.00\% & 62.80\% & 0.683 \\
2019+2018      & NaiveBayes           & 58.35\% & 58.40\% & 43.70\% & 58.20\% & 58.40\% & 58.30\% & 0.584 \\
2020           & J48                  & 79.95\% & 80.00\% & 20.30\% & 80.30\% & 80.00\% & 80.00\% & 0.832 \\
2020           & Random Forest        & 80.66\% & 80.70\% & 20.70\% & 80.60\% & 80.70\% & 80.70\% & 0.869 \\
2020           & AdaBoostM1           & \cellcolor[HTML]{EFEFEF}81.36\% & 81.40\% & 16.70\% & \cellcolor[HTML]{EFEFEF}82.90\% & \cellcolor[HTML]{EFEFEF}81.40\% & \cellcolor[HTML]{EFEFEF}81.50\% & \cellcolor[HTML]{EFEFEF}0.870 \\
2020           & SMO                  & 80.77\% & 80.80\% & 20.20\% & 80.90\% & 80.80\% & 80.80\% & 0.803 \\
2020           & MultilayerPerceptron & 80.07\% & 80.10\% & 21.50\% & 80.00\% & 80.10\% & 80.00\% & 0.868 \\
2020           & NaiveBayes           & 73.39\% & 73.40\% & 33.00\% & 73.70\% & 73.40\% & 72.30\% & 0.783 \\
2020+2019      & J48                  & 76.32\% & 76.30\% & 22.10\% & 77.80\% & 76.30\% & 76.50\% & 0.801 \\
2020+2019      & Random Forest        & 77.19\% & 77.20\% & 22.60\% & 77.70\% & 77.20\% & 77.30\% & 0.828 \\
2020+2019      & AdaBoostM1           & 76.80\% & 76.80\% & 20.30\% & 79.40\% & 76.80\% & 76.90\% & 0.818 \\
2020+2019      & SMO                  & 75.53\% & 75.50\% & 25.00\% & 75.80\% & 75.50\% & 75.60\% & 0.753 \\
2020+2019      & MultilayerPerceptron & 75.85\% & 75.80\% & 24.60\% & 76.10\% & 75.80\% & 75.90\% & 0.822 \\
2020+2019      & NaiveBayes           & 68.43\% & 68.40\% & 35.70\% & 68.00\% & 68.40\% & 67.90\% & 0.742 \\
2020+2019+2018 & J48                  & 76.40\% & 76.40\% & 22.70\% & 77.40\% & 76.40\% & 76.50\% & 0.792 \\
2020+2019+2018 & Random Forest        & 76.77\% & 76.80\% & 22.70\% & 77.50\% & 76.80\% & 76.90\% & 0.828 \\
2020+2019+2018 & AdaBoostM1           & 76.40\% & 76.40\% & 20.50\% & 79.30\% & 76.40\% & 76.50\% & 0.807 \\
2020+2019+2018 & SMO                  & 75.19\% & 75.20\% & 24.60\% & 75.80\% & 75.20\% & 75.30\% & 0.753 \\
2020+2019+2018 & MultilayerPerceptron & 76.92\% & 76.90\% & 22.50\% & 77.70\% & 76.90\% & 77.10\% & 0.822 \\
2020+2019+2018 & NaiveBayes           & 68.18\% & 68.20\% & 35.70\% & 67.80\% & 68.20\% & 67.70\% & 0.736 \\
\noalign{\smallskip}\hline
\end{tabular}
\end{adjustbox}
\end{table}

\subsubsection{Best machine learning model for code smell God Class}
\label{ResultsRQ2-GC}
Table \ref{table:ResultsGC} presents the results of God Class detection using the 10-Fold Cross-Validation technique and where the best values are marked. As we can see in table \ref{table:ResultsGC}, the model that presents the best value for the ROC is Naive Bayes with a value of 0.896. For the remaining four evaluation metrics, the Random Forest model presents the same values as the Naive Bayes. Thus, the Naive Bayes and Random Forest models present an accuracy value of 88.97\%, a precision value of 89.70\%, a recall value of 89.00\%, and an f-measure value of 88.70\%.

When we evaluate the models by the ROC value, we verify that, except for the J48 model, all the other five models have values higher than 0.857. For the remaining evaluation metrics all six models have: a) accuracy values higher or equal to 87.50\%, b) precision values higher or equal to 87.80\%, c) recall values higher or equal to 87.50\%, and d) f-measure values higher or equal to 87.20\%.   

When we compare the results of the code smell God Class detection with those of the Long Method, we verify that the results of God class are better.

\begin{table}[ht]
\centering
\caption{God Class: Performance of the code smell prediction models}
\label{table:ResultsGC}
\setlength{\tabcolsep}{4pt}
\begin{adjustbox}{width=\textwidth}
\begin{tabular}{clccccccc}
\hline\noalign{\smallskip}
\textbf{Dataset} & \multicolumn{1}{c}{\textbf{Classifier}} & \textbf{Accuracy} & \textbf{TP Rate} & \textbf{FP Rate} & \textbf{Precision} & \textbf{Recall} & \textbf{F-Measure} & \textbf{ROC Area} \\
\noalign{\smallskip}\hline\noalign{\smallskip}
2018           & J48                  & 81.82\% & 81.80\% & 26.50\% & 82.00\% & 81.80\% & 81.10\% & 0.692 \\
2018           & Random Forest        & 63.64\% & 63.60\% & 47.60\% & 61.90\% & 63.60\% & 62.30\% & 0.491 \\
2018           & AdaBoostM1           & 68.18\% & 68.20\% & 39.60\% & 67.40\% & 68.20\% & 67.70\% & 0.571 \\
2018           & SMO                  & 77.27\% & 77.30\% & 29.10\% & 76.90\% & 77.30\% & 76.90\% & 0.741 \\
2018           & MultilayerPerceptron & 72.73\% & 72.70\% & 31.70\% & 72.70\% & 72.70\% & 72.70\% & 0.768 \\
2018           & NaiveBayes           & 68.18\% & 68.20\% & 45.00\% & 66.70\% & 68.20\% & 66.10\% & 0.651 \\
2019           & J48                  & 72.87\% & 72.90\% & 29.50\% & 72.70\% & 72.90\% & 72.70\% & 0.725 \\
2019           & Random Forest        & 73.64\% & 73.60\% & 28.50\% & 73.50\% & 73.60\% & 73.50\% & 0.802 \\
2019           & AdaBoostM1           & 72.87\% & 72.90\% & 29.50\% & 72.70\% & 72.90\% & 72.70\% & 0.793 \\
2019           & SMO                  & 76.74\% & 76.70\% & 26.60\% & 76.90\% & 76.70\% & 76.30\% & 0.751 \\
2019           & MultilayerPerceptron & 75.97\% & 76.00\% & 27.70\% & 76.10\% & 76.00\% & 75.50\% & 0.797 \\
2019           & NaiveBayes           & 76.74\% & 76.70\% & 26.60\% & 76.90\% & 76.70\% & 76.30\% & 0.804 \\
2019+2018      & J48                  & 70.86\% & 70.90\% & 30.00\% & 70.80\% & 70.90\% & 70.80\% & 0.693 \\
2019+2018      & Random Forest        & 67.55\% & 67.50\% & 32.40\% & 67.80\% & 67.50\% & 67.60\% & 0.781 \\
2019+2018      & AdaBoostM1           & 69.54\% & 69.50\% & 30.90\% & 69.50\% & 69.50\% & 69.50\% & 0.771 \\
2019+2018      & SMO                  & 72.19\% & 72.20\% & 28.90\% & 72.10\% & 72.20\% & 72.00\% & 0.716 \\
2019+2018      & MultilayerPerceptron & 71.52\% & 71.50\% & 29.00\% & 71.50\% & 71.50\% & 71.50\% & 0.805 \\
2019+2018      & NaiveBayes           & 74.83\% & 74.80\% & 26.50\% & 74.90\% & 74.80\% & 74.60\% & 0.669 \\
2020           & J48                  & 87.50\% & 87.50\% & 17.30\% & 87.80\% & 87.50\% & 87.20\% & 0.791 \\
2020           & Random Forest        & \cellcolor[HTML]{EFEFEF}88.97\% & 89.00\% & 16.40\% & \cellcolor[HTML]{EFEFEF}89.70\% & \cellcolor[HTML]{EFEFEF}89.00\% & \cellcolor[HTML]{EFEFEF}88.70\% & 0.893 \\
2020           & AdaBoostM1           & 88.24\% & 88.20\% & 16.80\% & 88.70\% & 88.20\% & 87.90\% & 0.876 \\
2020           & SMO                  & 88.24\% & 88.20\% & 16.80\% & 88.70\% & 88.20\% & 87.90\% & 0.857 \\
2020           & MultilayerPerceptron & 88.24\% & 88.20\% & 16.80\% & 88.70\% & 88.20\% & 87.90\% & 0.885 \\
2020           & NaiveBayes           & \cellcolor[HTML]{EFEFEF}88.97\% & 89.00\% & 16.40\% & \cellcolor[HTML]{EFEFEF}89.70\% & \cellcolor[HTML]{EFEFEF}89.00\% & \cellcolor[HTML]{EFEFEF}88.70\% & \cellcolor[HTML]{EFEFEF}0.896 \\
2020+2019      & J48                  & 82.64\% & 82.60\% & 21.70\% & 82.90\% & 82.60\% & 82.30\% & 0.759 \\
2020+2019      & Random Forest        & 83.02\% & 83.00\% & 21.50\% & 83.40\% & 83.00\% & 82.60\% & 0.850 \\
2020+2019      & AdaBoostM1           & 82.64\% & 82.60\% & 21.70\% & 82.90\% & 82.60\% & 82.30\% & 0.857 \\
2020+2019      & SMO                  & 82.26\% & 82.30\% & 22.30\% & 82.60\% & 82.30\% & 81.90\% & 0.800 \\
2020+2019      & MultilayerPerceptron & 82.26\% & 82.30\% & 22.30\% & 82.60\% & 82.30\% & 81.90\% & 0.853 \\
2020+2019      & NaiveBayes           & 83.02\% & 83.00\% & 21.50\% & 83.40\% & 83.00\% & 82.60\% & 0.859 \\
2020+2019+2018 & J48                  & 81.88\% & 81.90\% & 21.70\% & 82.30\% & 81.90\% & 81.50\% & 0.763 \\
2020+2019+2018 & Random Forest        & 81.53\% & 81.50\% & 22.00\% & 81.90\% & 81.50\% & 81.20\% & 0.853 \\
2020+2019+2018 & AdaBoostM1           & 80.84\% & 80.80\% & 22.70\% & 81.20\% & 80.80\% & 80.50\% & 0.854 \\
2020+2019+2018 & SMO                  & 83.28\% & 83.30\% & 20.30\% & 83.80\% & 83.30\% & 82.90\% & 0.815 \\
2020+2019+2018 & MultilayerPerceptron & 82.23\% & 82.20\% & 20.10\% & 82.20\% & 82.20\% & 82.10\% & 0.880 \\
2020+2019+2018 & NaiveBayes           & 81.88\% & 81.90\% & 21.30\% & 82.10\% & 81.90\% & 81.60\% & 0.731 \\
\noalign{\smallskip}\hline
\end{tabular}
\end{adjustbox}
\end{table}

\subsubsection{Best machine learning model for code smell Feature Envy}
\label{ResultsRQ2-FE}
Regarding the code smell Feature Envy, we present in table \ref{table:ResultsFE} the results of the evaluation of the different models. As explained in point \ref{Data}, we do not have precision and f-measure results for the models created from the datasets 2018 and 2020, for this reason we will not consider in the response to the RQ the models resulting from the training by these two datasets.

When we evaluate the models by the ROC metric, we find that the best model is the Random Forrest with a ROC of 0.570. However, if we compare the various evaluation metrics we find that all the other evaluation metrics have better values than the ROC metric. The best performance in the detection of Feature Envy is obtained by the Naive Bayes model for precision with a value of 61.40\%. The Random Forrest model also obtains the best accuracy with 59.69\% and recall with a value of 59.70\%.

When we compare the results of the models for the detection of the three smells, we verify that the worst results are obtained by the Feature Envy detection models and the best results are obtained by the God Class detection models.

\begin{table}[ht]
\centering
\caption{Feature Envy: Performance of the code smell prediction models}
\label{table:ResultsFE}
\setlength{\tabcolsep}{4pt}
\begin{adjustbox}{width=\textwidth}
\begin{tabular}{clccccccc}
\hline\noalign{\smallskip}
\textbf{Dataset} & \multicolumn{1}{c}{\textbf{Classifier}} & \textbf{Accuracy} & \textbf{TP Rate} & \textbf{FP Rate} & \textbf{Precision} & \textbf{Recall} & \textbf{F-Measure} & \textbf{ROC Area} \\
\noalign{\smallskip}\hline\noalign{\smallskip}
2018             & J48                                     & 70.00\%           & 70.00\%          & 70.00\%          & -                  & 70.00\%         & -                  & 0.000             \\
2018             & Random Forest                           & 70.00\%           & 70.00\%          & 70.00\%          & -                  & 70.00\%         & -                  & 0.000             \\
2018             & AdaBoostM1                              & 70.00\%           & 70.00\%          & 70.00\%          & -                  & 70.00\%         & -                  & 0.000             \\
2018             & SMO                                     & 70.00\%           & 70.00\%          & 70.00\%          & -                  & 70.00\%         & -                  & 0.000             \\
2018             & MultilayerPerceptron                    & 70.00\%           & 70.00\%          & 70.00\%          & -                  & 70.00\%         & -                  & 0.000             \\
2018             & NaiveBayes                              & 30.00\%           & 30.00\%          & 87.10\%          & 35.00\%            & 30.00\%         & 32.30\%            & 0.000             \\
2019             & J48                                     & 56.85\%           & 56.90\%          & 46.10\%          & 56.20\%            & 56.90\%         & 56.20\%            & 0.563             \\
2019             & Random Forest                           & 58.38\%           & 58.40\%          & 44.70\%          & 57.80\%            & 58.40\%         & 57.70\%            & \cellcolor[HTML]{EFEFEF}0.570             \\
2019             & AdaBoostM1                              & 54.82\%           & 54.80\%          & 51.40\%          & 52.90\%            & 54.80\%         & 51.40\%            & 0.548             \\
2019             & SMO                                     & 52.79\%           & 52.80\%          & 51.30\%          & 51.50\%            & 52.80\%         & 51.40\%            & 0.508             \\
2019             & MultilayerPerceptron                    & 51.78\%           & 51.80\%          & 52.30\%          & 50.40\%            & 51.80\%         & 50.40\%            & 0.544             \\
2019             & NaiveBayes                              & 52.28\%           & 52.30\%          & 45.40\%          & 54.30\%            & 52.30\%         & 52.00\%            & 0.547             \\
2019+2018        & J48                                     & 57.97\%           & 58.00\%          & 42.80\%          & 57.90\%            & 58.00\%         & \cellcolor[HTML]{EFEFEF}58.00\%            & 0.552             \\
2019+2018        & Random Forest                           & 57.49\%           & 57.50\%          & 43.80\%          & 57.30\%            & 57.50\%         & 57.30\%            & 0.542             \\
2019+2018        & AdaBoostM1                              & 53.62\%           & 53.60\%          & 48.80\%          & 52.90\%            & 53.60\%         & 52.90\%            & 0.554             \\
2019+2018        & SMO                                     & 55.56\%           & 55.60\%          & 45.40\%          & 55.50\%            & 55.60\%         & 55.50\%            & 0.551             \\
2019+2018        & MultilayerPerceptron                    & 53.62\%           & 53.60\%          & 47.50\%          & 53.50\%            & 53.60\%         & 53.50\%            & 0.548             \\
2019+2018        & NaiveBayes                              & 51.69\%           & 51.70\%          & 47.30\%          & 52.60\%            & 51.70\%         & 51.70\%            & 0.548             \\
2020             & J48                                     & 64.23\%           & 64.20\%          & 64.20\%          & -                  & 64.20\%         & -                  & 0.467             \\
2020             & Random Forest                           & 64.23\%           & 64.20\%          & 64.20\%          & -                  & 64.20\%         & -                  & 0.486             \\
2020             & AdaBoostM1                              & 64.23\%           & 64.20\%          & 64.20\%          & -                  & 64.20\%         & -                  & 0.468             \\
2020             & SMO                                     & 64.23\%           & 64.20\%          & 64.20\%          & -                  & 64.20\%         & -                  & 0.500             \\
2020             & MultilayerPerceptron                    & 64.23\%           & 64.20\%          & 64.20\%          & -                  & 64.20\%         & -                  & 0.536             \\
2020             & NaiveBayes                              & 51.22\%           & 51.20\%          & 38.20\%          & \cellcolor[HTML]{EFEFEF}61.40\%            & 51.20\%         & 50.90\%            & 0.482             \\
2020+2019        & J48                                     & 59.38\%           & 59.40\%          & 56.70\%          & 57.00\%            & 59.40\%         & 48.40\%            & 0.529             \\
2020+2019        & Random Forest                           & \cellcolor[HTML]{EFEFEF}59.69\%           & 59.70\%          & 56.10\%          & 58.00\%            & \cellcolor[HTML]{EFEFEF}59.70\%         & 49.40\%            & 0.548             \\
2020+2019        & AdaBoostM1                              & 58.75\%           & 58.80\%          & 59.30\%          & 34.80\%            & 58.80\%         & 43.70\%            & 0.519             \\
2020+2019        & SMO                                     & 59.06\%           & 59.10\%          & 56.50\%          & 55.70\%            & 59.10\%         & 49.10\%            & 0.513             \\
2020+2019        & MultilayerPerceptron                    & 57.50\%           & 57.50\%          & 56.40\%          & 52.80\%            & 57.50\%         & 49.80\%            & 0.545             \\
2020+2019        & NaiveBayes                              & 52.81\%           & 52.80\%          & 40.70\%          & 58.70\%            & 52.80\%         & 51.90\%            & 0.532             \\
2020+2019+2018   & J48                                     & 57.58\%           & 57.60\%          & 57.80\%          & 50.50\%            & 57.60\%         & 44.50\%            & 0.518             \\
2020+2019+2018   & Random Forest                           & 58.48\%           & 58.50\%          & 55.90\%          & 56.20\%            & 58.50\%         & 47.70\%            & 0.539             \\
2020+2019+2018   & AdaBoostM1                              & 57.88\%           & 57.90\%          & 58.40\%          & 33.80\%            & 57.90\%         & 42.70\%            & 0.498             \\
2020+2019+2018   & SMO                                     & 58.79\%           & 58.80\%          & 54.70\%          & 56.90\%            & 58.80\%         & 49.70\%            & 0.520             \\
2020+2019+2018   & MultilayerPerceptron                    & 54.85\%           & 54.80\%          & 58.50\%          & 46.80\%            & 54.80\%         & 45.50\%            & 0.533             \\
2020+2019+2018   & NaiveBayes                              & 51.82\%           & 51.80\%          & 43.20\%          & 56.10\%            & 51.80\%         & 51.10\%            & 0.524 \\           
\noalign{\smallskip}\hline
\end{tabular}
\end{adjustbox}
\end{table}

\subsection{\textbf{RQ3. It is possible to use Collective Knowledge for code smells detection?}}
\label{ResultsRQ3}
Several studies present code smells detection results through machine learning techniques with accuracy, precision, recall, and f-measure, very close to 100\%. However, these studies use very treated datasets to obtain good results, which makes the datasets unrealistic. A proof of this is the replication of one of the most important studies on code smells detection using machine learning techniques by \cite{Nucci2018}, where more realistic datasets were used in this replication. The results of this replication show that the accuracy value, on average, decreased from 96\% to 76\%, but the f-measure presented results 90\% lower than in the reference work. When we compare our results with the \cite{Nucci2018} study, we find that the results are similar in some metrics, and better in others.

As reported in the answers to RQ1 and RQ2, we obtained values for some machine learning models close to 90\%, which can be considered very good. The fact that the most recent datasets are the ones that usually present the best results, mainly the year 2020, leaves us with expectations of being able to improve the results further. This improvement is mainly due to the improvement of the methodological process, which has been progressively refined each year. 
Thus, the answer to this RQ is yes, it is possible to use Collective Knowledge for code smells detection.


\section{Threats to validity}
\label{sec:Threatsvalidity}
In our study, we made assumptions that may threaten the validity of our results. In this section, we discuss possible sources of threats, and how we mitigated them.

\subsection{Conclusion Validity}
\label{ConclusionValidity}
Threats in this category impact the relation between treatment and outcome. 

The first threat is in the evaluation methodology, so we adopted the 10-fold cross-validation, which is one of the most used in machine learning, and to directly compare our results with those achieved in the other study's. 

As for the evaluation metrics adopted to interpret the performance of the experimented models, we have adopted the most common machine learning metrics, which have been used in other studies with some similarity.  

To test if there was a statistically significant difference between the performance of the six machine learning models, we used the one-way analysis of variance (ANOVA).

\subsection{Construct Validity}
\label{ConstructValidity}
As for potential issues related to the relationship between theory and observation, we may have been subject of problems in the adopted methodology.
To avoid bias in process, We elaborated a script in which we detailed all the steps that the teams had to carry out to detect code smells, so that there would be uniformity in the process. However, we cannot guarantee the correct use of this script by all the teams.

Code metrics are extremely important because they play the role of independent variables in the machine learning algorithms. To avoid bias in metrics extraction we used the same metrics as in \cite{Fontana2015}, since they are publicly available.
As for the experimented prediction models, we exploited the implementation provided by the Weka framework \citep{Hall2009}, which is widely considered as a reliable tool. To avoid bias in the parameterization of the Weka algorithms, we used the default values for the parameters.

\subsection{Internal Validity}
\label{InternalValidity}
This threat is related to the correctness of the experiments’ outcome. Since the definition of code smells is subjective, it may cause different interpretations and, as such, the manual evaluation is not entirely reliable. To mitigate this problem, an advisor is used in the experiment to serve as a basis for identifying code smells, although the final decision is always made by the team, which is composed of several developers, and all had the same training.

The maturity, experience, and knowledge of team members about code smells is a variable that we cannot control. As such, there may be variations in the accuracy and precision of the detection of code smells. To minimize the possible bias, the decisions are not individual, but taken by the team.

Because code smells is only detected in three Java projects, there may be some bias as to the number and type of CS existing in these Java projects. We chose these projects because they are open-source, are widely used in CS detection, and are not toy examples due to their considerable dimension. 

\subsection{External Validity}
\label{ExternalValidity}
Finally, the External validity is concerned with whether we can generalize the results outside the scope of our study. 

With respect to the generalizability, we used the three most common code smells in this type of studies. Regarding the code metrics, we used a high number, 61 metrics for God Class and 82 metrics for Feature Envy and Long Method, thus ensuring a wide scope. 

In terms of programming language, we only used Java projects, but Java is one of the most used languages in code smells detection studies. According to our SLR \cite{Reis2020}, we found that seven languages were used in detection studies, Java is the first, accounting for 77.1\% of the studies.


\section{Conclusion and future work}
\label{sec:Conclusion}
We propose the concept of crowdsmelling \citep{Reis2017} – use of collective intelligence in the detection of code smells – to mitigate the aforesaid problems of subjectivity and lack of calibration data required to obtain accurate detection model parameters.
In this paper we reported first results of a study investigating the approach CrowdSmelling, a collaborative crowdsourcing approach, based in machine learning, where the wisdom of the crowd (of software developers) will be used to collectively calibrate code smells detection algorithms. 

For 3 years we collected CS detection data by several teams manually, although they could use JDeodorant as an advisor if they wanted.  Combining the data from each year with the previous ones, we created several oracles for each of the three code smells (Long Method, God Class, Feature Envy). The latter were used to train a set of machine learning algorithms, creating the detection models for each of the three code smells, in a total of 108 models. Finally, to evaluate the models we tested them using the 10-Fold Cross-Validation methodology, and analyzed the metrics Accuracy, Precision, Recall, and F-Measure, with special emphasis on ROC, because the datasets were not treated, for example, balanced. This way we created the most realistic datasets possible. To check if there were significant differences between the classifications presented by the different models, we proceeded to the analysis of variance through a one-way analysis of variance (ANOVA).

Regarding RQ1, we conclude that the best results for the code smell Long Method were obtained by the Random Forrest and AdaBoostM1 algorithms. The best result with a ROC of 0.870 was obtained by AdaBoostM1 when trained by the dataset 2020, followed by the Random Forrest with ROC of 0.869 for the same dataset. For the code smell God Class, the best result was obtained by the NaiveBayes algorithm, when trained by the dataset 2020, with the ROC value of 0.896. For Feature Envy the results are low, with the Random Forest algorithm having the best ROC value of 0.570 when trained by dataset 2019.
The results of the variance tests (ANOVA) show there was no statistically significant difference between the performance of the six machine learning models, when trained with data from the crowd and therefore more realistic.

As for RQ2, the best machine learning model for Long Method detection is AdaBoostM1, presenting the best values for all evaluation metrics, a ROC value of 0.870, a accuracy of 81.36\%, a precision of 82.90\%, a recall of 81.40\%, and F-measure of 81.50\%. For the God Class, the model that presents the best value for the ROC is Naive Bayes with a value of 0.896. the Naive Bayes and Random Forest models present an accuracy value of 88.97\%, a precision value of 89.70\%, a recall value of 89.00\%, and an f-measure value of 88.70\%. For the Feature Envy, best model is the Random Forrest with a ROC of 0.570. However, the best performance in the detection of Feature Envy is obtained by the Naive Bayes model for precision with a value of 61.40\%.   

Regarding RQ3, it is possible to use the crowdsmelling – use of collective intelligence in the detection of code smells – as a good approach to detection of code smells, because we obtained values for some machine learning models close to 90\%, which can be considered very good, for realistic datasets, which reflect the detection performed by developers. 
The fact that the most recent datasets, the year 2020, are the ones that usually presented the best results, leaves us with great motivation to continue developing this detection approach because we think that we can even better the results.  

We are currently developing a plugin for the Eclipse IDE, which extracts the code metrics, detects the code smells, identifies the code smells in the code, receives the programmer's opinion regarding the detection of the code smell (i.e., if the programmer agrees or not with the code smell and stores all this information in a database). This plugin is expected to simplify the use of the crowdsmelling approach, making it simple for programmers to use when developing their Java projects.   

\begin{acknowledgements}
This work was partially funded by the Portuguese Foundation for Science and Technology, under ISTAR's projects UIDB/04466/2020 and UIDP/04466/2020.
\end{acknowledgements}

%
%

\bibliographystyle{apalike}
\bibliography{Crowd}

\appendix
\section*{\large{Appendices}}
\addcontentsline{toc}{section}{Appendices}
\renewcommand{\thesubsection}{Appendix \Alph{subsection}.}

\subsection{Code metrics}
\label{AppendixA}

\footnotesize{
\begin{tabular}{ll}
\hline\noalign{\smallskip}
Metric & Acronym \\
\noalign{\smallskip}\hline\noalign{\smallskip}
Lines   of Code                                             & LOC       \\
Lines   of Code Without Accessor or Mutator Methods         & LOCNAMM   \\
Number   of Packages                                        & NOPK      \\
Number   of Classes                                         & NOCS      \\
Number   of Methods                                         & NOM       \\
Number   of Not Accessor or Mutator Methods                 & NOMNAMM   \\
Number   of Attributes                                      & NOA       \\
Cyclomatic   Complexity                                     & CYCLO     \\
Weighted   Methods Count                                    & WMC       \\
Weighted   Methods Count of Not Accessor or Mutator Methods & WMCNAMM   \\
Average   Methods Weight                                    & AMW       \\
Average   Methods Weight of Not Accessor or Mutator Methods & AMWNAMM   \\
Maximum   Nesting Level                                     & MAXNESTING\\
Weight   of Class                                           & WOC       \\
Called   Local Not Accessor or Mutator Methods              & CLNAMM    \\
Number   of Parameters                                      & NOP       \\
Number   of Accessed Variables                              & NOAV      \\
Access   to Local Data                                      & ATLD      \\
Number   of Local Variable                                  & NOLV      \\
Tight   Class Cohesion                                      & TCC       \\
Lack   of Cohesion in Methods                               & LCOM      \\
Fanout                                                      & FANOUT    \\
Access   to Foreign Data                                    & ATFD      \\
Foreign   Data Providers                                    & FDP       \\
Response   for A Class                                      & RFC       \\
Coupling   Between Objects Classes                          & CBO       \\
Called   Foreign Not Accessor or Mutator Methods            & CFNAMM    \\
Coupling   Intensity                                        & CINT      \\
Coupling   Dispersion                                       & CDISP     \\
Maximum   Message Chain Length                              & MAMCL     \\
Number   of Message Chain Statements                        & NMCS      \\
Mean   Message Chain Length                                 & MEMCL     \\
Changing   Classes                                          & CC        \\
Changing   Methods                                          & CM        \\
Number   of Accessor Methods                                & NOAM      \\
Number   of Public Attributes                               & NOPA      \\
Locality   of Attribute Accesses                            & LAA       \\
Depth   of Inheritance Tree                                 & DIT       \\
Number   of Interfaces                                      & NOI       \\
Number   of Children                                        & NOC       \\
Number   of Methods Overridden                              & NMO       \\
Number   of Inherited Methods                               & NIM       \\
Number   of Implemented Interfaces                          & NOII      \\
Number   of Default Attributes                              & NODA      \\
Number   of Private Attributes                              & NOPVA     \\
Number   of Protected Attributes                            & NOPRA     \\
Number   of Final Attributes                                & NOFA      \\
Number   of Final and Static Attributes                     & NOFSA     \\
Number   of Final and Non - Static Attributes               & NOFNSA    \\
Number   of Not Final and Non - Static Attributes           & NONFNSA   \\
Number   of Static Attributes                               & NOSA      \\
Number   of Non - Final and Static Attributes               & NONFSA    \\
Number   of Abstract Methods                                & NOABM     \\
Number   of Constructor Methods                             & NOCM      \\
Number   of Non - Constructor Methods                       & NONCM     \\
Number   of Final Methods                                   & NOFM      \\
Number   of Final and Non - Static Methods                  & NOFNSM    \\
Number   of Final and Static Methods                        & NOFSM     \\
Number   of Non - Final and Non - Abstract Methods          & NONFNABM  \\
Number   of Final and Non - Static Methods                  & NONFNSM   \\
Number   of Non - Final and Static Methods                  & NONFSM    \\
Number   of Default Methods                                 & NODM      \\
Number   of Private Methods                                 & NOPM      \\
Number   of Protected Methods                               & NOPRM     \\
Number   of Public Methods                                  & NOPLM     \\
Number   of Non - Accessors Methods                         & NONAM     \\
Number   of Static Methods                                  & NOSM      \\                
\noalign{\smallskip}\hline
\end{tabular}
}

\end{document}